\documentclass{moriond}

\def\be{\begin{equation}}
\def\ee{\end{equation}}
\def\bea{\begin{eqnarray}}
\def\eea{\end{eqnarray}}

\newcommand{\Photo}{}

\usepackage{lineno}

\begin{document}
\vspace*{4cm}
\title{Standard Model Soft and Hard QCD at ATLAS and CMS}

\author{M. Trzebi\'nski\\ on behalf of ATLAS and CMS Collaborations}

\address{The Henryk Niewodnicza\'nski Institute of Nuclear Physics Polish Academy of Sciences,\\
Radzikowskiego 152, 31-342 Krak\'ow, Poland}

\maketitle\abstracts{
Quantum Chromodynamics (QCD) is explored at the Large Hadron Collider (LHC) over a wide range of energy scales, from soft hadronic interactions to hard parton scattering. 
This contribution presents a selection of recent measurements from the ATLAS and CMS experiments probing QCD in proton--proton collisions at the LHC. 
The results span a broad range of momentum scales, from studies of charged-particle event-shape observables in minimum-bias events, which are sensitive to soft QCD dynamics and hadronisation, to measurements of inclusive di-jet production that provide stringent tests of perturbative QCD. 
Recent studies of double-parton scattering in same-sign $WW$ production are also discussed, offering new insight into multi-parton interactions inside the proton. 
Finally, the capabilities of the forward proton detectors operated by the ATLAS and CMS collaborations are briefly reviewed, highlighting their contribution to the LHC physics programme through measurements of diffractive and photon-induced processes.
}

\section{Introduction}

The Large Hadron Collider (LHC) \cite{LHC} provides a unique environment for studying Quantum Chromodynamics (QCD) \cite{QCDreview} over a broad range of momentum scales. 
Measurements of soft hadronic activity probe the non-perturbative regime of strong interactions, including hadronisation and the underlying event. 
Hard scattering processes test perturbative QCD (pQCD) calculations at the highest currently accessible energies. 
Together, these studies provide stringent tests of theoretical predictions and Monte Carlo (MC) event generators and improve the modelling required for precision measurements and searches for physics Beyond the Standard Model (BSM).

The ATLAS \cite{ATLAS} and CMS \cite{CMS} experiments continue to deliver a broad programme of QCD measurements using proton--proton collision data collected during LHC Run~2 (2015--2018) at $\sqrt{s}=13$~TeV. 
Moreover, the increasing data sample accumulated during the LHC Run~3 provides unprecedented opportunities to extend previous studies into new kinematic regimes and significantly improve the precision of QCD measurements.

\section{Soft QCD: Event Shapes in Minimum-bias Events}

Measurements of event-shape observables in minimum-bias proton--proton collisions provide a sensitive probe of soft QCD dynamics, including hadronisation and multi-parton interactions. 
These observables characterise the global geometry of the charged-particle production and are particularly useful for testing and constraining phenomenological models implemented in Monte Carlo event generators.

The CMS Collaboration has recently measured the charged-particle event-shape variables~\cite{CMS_Event_Shapes}. 
This analysis used proton--proton collision data collected at $\sqrt{s}=13$~TeV in a special run with approximately one inelastic interaction per bunch crossing: pile-up $\mu = 1$. 
The data correspond to an integrated luminosity of 64~$\mu$b$^{-1}$. 
The analysis examined a set of global event-shape observables, including multiplicity, total invariant mass, sphericity, thrust, transverse thrust, broadening, transverse sphericity, isotropy, and related quantities, characterising the overall topology of the hadronic final state.

The measured distributions, as well as their dependence on the charged-particle multiplicity, were unfolded to the particle level.
Predictions of several Monte Carlo event generators were compared to the experimental data. 
None of the considered models provides a satisfactory simultaneous description of all observables, with common features indicating that the data are systematically more isotropic than the simulations~\cite{CMS_Event_Shapes}. 
This behaviour highlights persistent deficiencies in the modelling of soft particle production and multi-parton interactions.

\section{Hard QCD: Jet Production}
Jets provide a direct experimental access to the dynamics of parton showering, hadronisation and the interplay between perturbative and non-perturbative QCD effects.

\subsection{Inclusive Di-jet Cross-sections}
Inclusive di-jet production at the LHC provides one of the most precise and robust tests of pQCD. 
Owing to the large production cross-section, di-jet events can be studied over a wide kinematic range, probing parton dynamics at momentum transfers extending into the multi-TeV regime. 
These measurements are sensitive to parton distribution functions and the strong coupling constant and constitute an essential benchmark for precision SM predictions.

Recent measurements by the ATLAS Collaboration use 140~fb$^{-1}$ of proton--proton collision data collected at $\sqrt{s}=13$~TeV~\cite{ATLAS_dijet}. 
The CMS analysis is based on 24.7~pb$^{-1}$ of data collected at $\sqrt{s}=5.02$~TeV \cite{CMS_dijet}. 
In both cases, the double-differential inclusive di-jet production cross-sections are measured as a function of the di-jet invariant mass and rapidity separation. 
The ATLAS measurement extends into the multi-TeV di-jet mass region, $m_{jj}$, while the lower-energy CMS result provides complementary constraints at smaller momentum scales. 
The results are corrected for detector effects using unfolding techniques, enabling direct comparison with particle-level theoretical predictions.

The measured cross-sections are generally well described by perturbative QCD calculations at next-to-leading (NLO) and next-to-next-to-leading (NNLO) orders, supplemented by electroweak corrections. 
However, both measurements highlight important aspects of the theoretical uncertainties.

In the ATLAS measurement, a non-trivial dependence of the data-to-theory ratio on the di-jet invariant mass and rapidity separation, $y^*$, is observed. 
At low values of $y^*$, theoretical predictions tend to overestimate the data, whereas at higher rapidities, the deviations become more pronounced as $m_{jj}$ increases. 
These effects may indicate limitations of fixed-order calculations in certain regions of phase space.

The CMS study shows improved agreement between data and theory when renormalisation and factorisation scales are set to $\mu_R = \mu_F = H_T$ and demonstrates a strong dependence of theoretical uncertainties on the scale choice. 
While NNLO calculations generally reduce scale uncertainties at high transverse momentum, an increase is observed in some low-$p_T$ regions, reflecting residual theoretical ambiguities. 
The use of modern PDF sets further improves the stability of the predictions.

\subsection{Upsilon Fragmentation in Jets}
The production of heavy quarkonia inside a jet offers a sensitive probe of the interplay between perturbative heavy-quark production and non-perturbative bound-state formation. 
The CMS Collaboration has recently studied $\Upsilon$(nS) production as a function of jet observables, investigating how bottomonium states are distributed within the reconstructed jets in proton--proton collisions at $\sqrt{s}=13$~TeV \cite{CMS_upsilon}.

The $\Upsilon(1S)$, $\Upsilon(2S)$, and $\Upsilon(3S)$ in jets were measured. 
Predictions of the MC event generators incorporating various quarkonium production and fragmentation models were compared to the data.
The results provide constraints on the modelling of heavy-quark fragmentation into bound states and help to discriminate between different production mechanisms implemented in theoretical calculations, adding information to the ``quarkonia polarisation puzzle''.

\subsection{Dead-cone Effect in Jets}
The dead-cone effect is a fundamental prediction of Quantum Chromodynamics, describing the suppression of collinear gluon radiation from a massive quark within an angular region defined approximately by quark mass and energy: $\theta_0 = m_q / E_q$. 
This effect modifies the pattern of parton shower evolution for heavy-flavour quarks compared to light quarks.

The CMS Collaboration has studied this phenomenon using $t\bar{t}$ events at $\sqrt{s} = 13$~TeV and 59.8~fb$^{-1}$ of data \cite{CMS_deadcone}. 
The analysis focuses on $b$-quark jets and measures the differential emission density as a function of an angular variable related to the jet radius and the emission angle, $dN_{\mathrm{emission}}/d\ln(R/\Delta R)$.

The measured distribution exhibits a clear suppression of collinear radiation compared with predictions that incorporate a massless parton shower. 
In contrast, simulations including mass effects provide a significantly improved description of the data. 
A light-quark baseline shows a substantially stronger emission rate at small angles, further highlighting the role of the quark mass in shaping the radiation pattern.

\section{Double-parton Scattering in the Same-sign $WW$ Production}
At the LHC collision conditions, multiple parton interactions may occur within a single collision. 
In the case of double-parton scattering (DPS), two hard interactions take place independently in the same proton--proton interaction. 
Measurements of DPS provide valuable information on the spatial and momentum correlations among partons inside the proton. They are important for improving the modelling of multi-parton interactions in MC event generators.

The ATLAS Collaboration has recently reported a measurement of DPS in same-sign $WW$ production using the full Run~2 data set collected at $\sqrt{s}=13$~TeV~\cite{ATLAS_DPS_WW}. 
Same-sign $WW$ production offers a particularly clean experimental signature, with two isolated leptons of the same electric charge, missing transverse momentum from neutrinos, and a relatively small contribution from single-parton scattering processes after dedicated event selections. 
A multivariate analysis is employed to separate the DPS signal from the remaining backgrounds.

The DPS contribution is extracted from a fit to the data, allowing the determination of the effective cross-section $\sigma_{\mathrm{eff}} = 10.6 \pm 1.8~\mathrm{mb}$. 
The result is consistent within uncertainties with previous measurements in different final states.

\section{Forward proton detectors}
\
The physics programme of the ATLAS and CMS experiments is complemented by dedicated forward proton detectors, AFP~\cite{AFP} and PPS~\cite{PPS}, respectively. 
These systems measure protons scattered at very small angles with respect to the beam direction. 
By combining the proton tagging with information from the central detectors, these detectors enable the study of diffractive and photon-induced processes that cannot be fully reconstructed using the central detectors alone.

During Run~2 and especially Run~3, AFP and PPS have significantly extended the LHC physics programme. 
Tagged forward protons allow an investigation of hard diffraction and central exclusive production, providing unique information on the colour-singlet exchange and proton structure. 
Studies of photon-induced processes may, in addition to SM tests, open doors to BSM searches. 
Proton tagging offers a powerful background suppression for rare exclusive final states, thereby increasing the sensitivity of such analyses. 
As the majority of proton-tagged data was collected during LHC Run 3, the analysis of these data has just started and will continue for several years.

\section{Summary and outlook}

The ATLAS and CMS experiments continue to deliver a broad and diverse programme of QCD measurements at the LHC. 
The recent results illustrate the wide range of physics accessible at the energy frontier, from studies of soft hadronic activity and event topology, through precision measurements of jet production, to investigations of multi-parton interactions and forward proton tagging.

The increased centre-of-mass energy and enormous data set collected during Run~3 enable measurements with improved precision and extended kinematic reach. 
This promises even more stringent tests of perturbative and non-perturbative QCD, improved modelling of proton--proton collisions, and deepening our understanding of the structure of the proton. 
In particular, data from forward detectors should provide further input for precision Standard Model measurements and allow new types of searches for BSM phenomena.

\section*{Acknowledgments}

The work of MT was partially supported by the Polish National Science Centre (project no. \textit{UMO-2019/34/E/ST2/00393}). Copyright 2026 CERN for the benefit of the ATLAS and CMS Collaborations. \textit{CC-BY-4.0} license.

\section*{References}

\end{document}